\documentclass{aa}  
\usepackage{graphicx,natbib}
\usepackage{hyperref} 
\usepackage{txfonts}

\begin{document}
   \title{The nature of optical and near-infrared variability of \object{BL Lacertae}
   \thanks{Based on data taken and assembled by the WEBT collaboration and stored in the WEBT archive at the Osservatorio Astronomico di Torino - INAF (http://www.oato.inaf.it/blazars/webt/).}} 

   \author{V.~M.~Larionov              \inst{ 1,2}
   \and   M.~Villata                 \inst{ 3}
   \and   C.~M.~Raiteri               \inst{ 3}
   }

   \offprints{V.\ Larionov, {\tt vlar@astro.spbu.ru}}

   \institute{
Astronomical Institute, St.-Petersburg State University, Russia                                     
   \and   Isaac Newton Institute of Chile, St.-Petersburg Branch, Russia                                      
    \and      INAF, Osservatorio Astronomico di Torino, Italy  }

   \date{}

 
  \abstract
   {Since 1997, BL~Lacertae has undergone a phase of high optical activity,
with the occurrence of several prominent outbursts. 
Starting from 1999, the Whole Earth Blazar Telescope (WEBT) consortium has organized
various multifrequency campaigns on this blazar, collecting tens of thousands of data points.
One of the main issues in the analysis of this huge dataset has been the study of colour variability.}
   {The massive amount of optical and near-infrared data collected during the campaigns
 enables us to perform a deep analysis of multiband data, with the aim of understanding the
 flux variability mechanisms.}
   {We use a new approach for the analysis of these data, 
focusing on the source spectral evolution.}
   {We show that the overall behaviour of the BL~Lacertae light and colour curves 
can be explained in terms of changing viewing angle of a moving, discrete emitting region, which causes variable Doppler boosting of the corresponding radiation.}
   {A fractal helical structure is suggested to be at the origin of the different time scales 
of variability.}

   \keywords{galaxies: active -- 
             galaxies: BL Lacertae objects: general -- 
             galaxies: BL Lacertae objects: individual: \object{BL Lacertae} -- 
             galaxies: jets -- 
             galaxies: quasars: general}
   \titlerunning{The nature of \object{BL Lacertae} variability}
   \authorrunning{Larionov, Villata \& Raiteri}            
   \maketitle
\section{Introduction}

BL Lacertae ($z=0.0688 \pm 0.0002$; \citealt{Miller1977}) is the prototype of a class of active galactic nuclei (AGNs), the BL Lac objects, which are well known for their pronounced variability at all wavelengths, from the radio to the $\gamma$-ray band (see e.g.\ \citealt{Villata2002} for an extended description of this object). It has been one of the favourite targets of the Whole Earth Blazar Telescope 
(WEBT)\footnote{{\tt http://www.oato.inaf.it/blazars/webt/}}, which has organized several multifrequency campaigns on this source. 

The campaign carried out in May 2000 -- January 2001 was particularly successful, with periods of exceptionally dense sampling that allowed \citet{Villata2002} to follow the intranight flux variations in detail. The slope of the optical spectrum was found to be weakly sensitive to the long-term brightness trend, but to strictly follow the short-term flux variations, becoming bluer when brighter. The authors suggested that the
nearly achromatic modulation of the flux base level on long time scales is due to a variation of the relativistic Doppler beaming factor, and that this variation is likely due to a change of the viewing angle. 

A subsequent campaign was carried out in 2001--2003. On that occasion, historical optical and radio light curves were reconstructed back to 1968, and both colour and time series analyses were performed. 
Colour analysis on a longer time period confirmed the conclusions by \citet{Villata2002}, and allowed \citet{Villata2004a} to quantify the degree of chromatism of both the short-term and long-term variability components. 
The optical spectral behaviour was further analysed by \citet{Papadakis2007}, who interpreted 
the bluer-when-brighter mild chromatism of the long-term
variations in terms of Doppler factor variations due to changes in the
viewing angle of a curved and inhomogeneous jet.
The study of the optical and radio flux behaviour in the last forty years led \citet{Villata2009a}
to propose a more complex scenario, where the emitting plasma is flowing along a rotating helical path in a curved jet.
Indeed, \citet{Raiteri2009} claimed that synchrotron plus self-Compton emission from a helical jet can explain the broad-band spectral energy distribution (SED) of BL Lacertae, from the radio to the $\gamma$-ray band.

We notice that in all the above papers dealing with colour analysis, the host galaxy contribution was subtracted from the total flux densities to analyse the colour behaviour of the AGN alone. 

A different approach was adopted by Hagen-Thorn and coauthors \citep[see, e.g.][]{Hagen-Thorn2004}. 
It is based on the assumption that the flux changes within some time interval are due to a single variable source. 
If the variability is caused only by its flux variations and the relative SED remains unchanged, then in the {\it n}-dimensional flux density space
$\{F_1, ..., F_n\}$ ({\it n} is the number of spectral bands used in multicolour observations) 
the observational points must lie on straight lines. The slopes of these lines are the flux density ratios for
the various pairs of bands.
With some limitations, the opposite is also true: a linear relation
between observed flux densities at two different wavelengths during some period of flux variability
implies that the slope (flux ratio) does not change. Such a relation for several bands would indicate that the relative SED of the variable source remains steady and can be derived from the slopes of the lines.

That kind of analysis, using $BVRIJHK$ photometry obtained during 1999--2001, led \citet{Hagen-Thorn2004} to the conclusion that the spectrum of the variable source in BL~Lacertae can be described as $F_\nu \propto \nu^{-\alpha}$, with different values of $\alpha$ for the optical and NIR bands: $\alpha_{\rm opt}=1.25$ and $\alpha_{\rm NIR}=0.90$. The authors point out that this data set does not allow to discriminate whether a single variable source acts over the whole wavelength range.

The lack of spectral variability found by \citet{Hagen-Thorn2004} appears in contradiction with the bluer-when-brighter trend obtained in the previously mentioned works based on the WEBT data. We notice that a spectral flattening with brightness increase had already been found by \citet{Webb1998} when analysing $BVRI$ observations of BL Lacertae around its 1997 outburst.
These authors suggested three possible explanations for the source spectral variability, involving changes in the electron energy distribution, its turnover frequency, or the presence of multiple synchrotron contributions, but they were not able to favour one of them because of the lack of information at other wavelengths. A bluer-when-brighter behaviour was also detected by other authors in datasets more or less extended in time \citep[e.g.][]{Zhang2004,Stalin2006,Gu2006}.

In this paper we investigate the matter with a larger dataset, to shed light on the source spectral variability and its possible origin.
Our dataset is presented in Sect.~2, where light curves and flux-flux plots are shown and spectral variability discussed. In Sect.~3 we address the question of the origin of the variability, investigating the possibility that it is due to changes of the Doppler factor, which in turn are caused by orientation effects. Sect.~4 contains a summary of the results.

\section{Observational data and analysis}

One of the main goals of all the multiwavelength campaigns carried on during the last decades was the investigation of colour variability on different portions of the light curve, as well as the determination of correlations and temporal delays between variations at different frequencies. 
Apparently, the former issue can be easily addressed by acquiring good-quality photometric data at different brightness levels, and then calculating the corresponding colour indices. However, the interpretation of the colour behaviour of blazars, particularly of BL~Lacertae, still remains a topic of discussion by several teams of observers. It may be, at least partly, explained by the complexity of the source itself, since one needs to subtract the radiation of the underlying elliptical galaxy, which is bright enough to contaminate the active nucleus photometry, especially at low flux levels. 

\begin{figure}
   \centering
  \resizebox{\hsize}{!}{\includegraphics[clip]{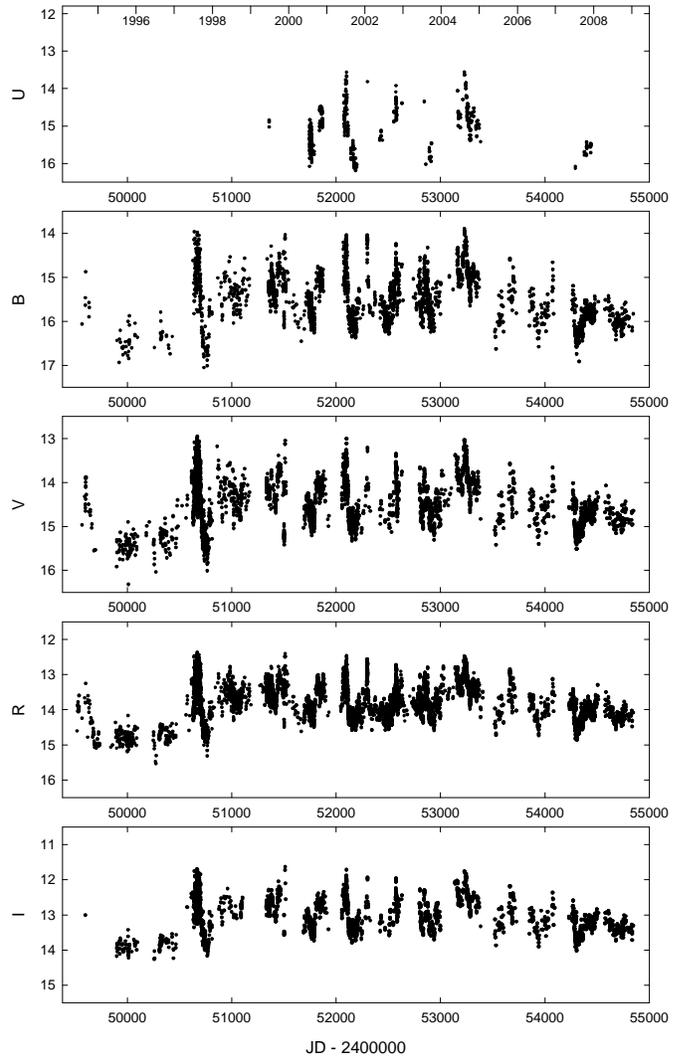}}
        \caption{Optical light curves of BL Lacertae in 1994--2008.} 
         \label{bl97_08opt}
   \end{figure}

\begin{figure}
   \centering
  \resizebox{\hsize}{!}{\includegraphics[clip]{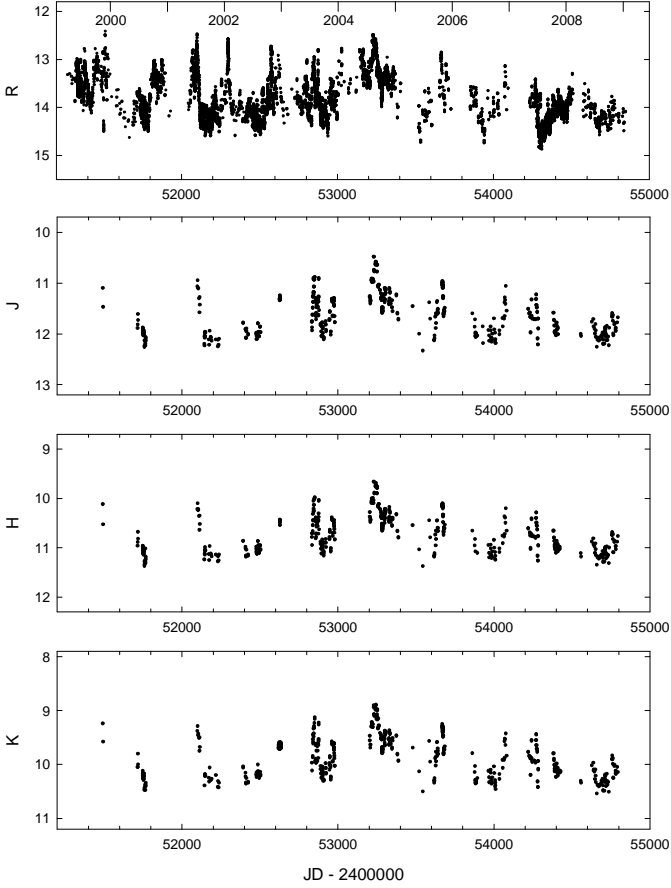}}
      \caption{Near-infrared and $R$-band light curves of BL Lacertae in 1999--2008.}
         \label{bl99_08ir}
   \end{figure}

The optical and near-infrared light curves discussed in the following are shown in Figs.~\ref{bl97_08opt} and~\ref{bl99_08ir}. They include all the WEBT data, complemented by Crimean and St.\ Petersburg optical data, and Campo Imperatore near-infrared data, acquired in 2005--2006 and 2008, i.e.\ in the periods not covered by WEBT campaigns.

The WEBT data had already been carefully assembled and ``cleaned''. The colour analysis needs even more attention. We included for further processing only pairs of data that (1) were obtained with the same instrumentation within 15 minutes from each other; (2) did not show systematic differences from the most reliable data sets; (3) did not present scatter caused by low intrinsic accuracy. After this screening, we were left with the following numbers of data pairs for each colour: 63 ($U-R$), 460 ($B-R$), 1006 ($V-R$), 1004 ($I-R$), 300 ($J-R$), 490 ($H-J$), and 473 ($K-J$).

We transformed magnitudes into de-reddened flux densities using the Galactic absorption $A_B=1\fm42$ from  \citet{Schlegel1998}, the \citet{Cardelli1989} extinction law, and the \citet{Bessell1998} flux-magnitude calibrations.

\begin{figure}
   \centering
  \resizebox{\hsize}{!}{\includegraphics[clip]{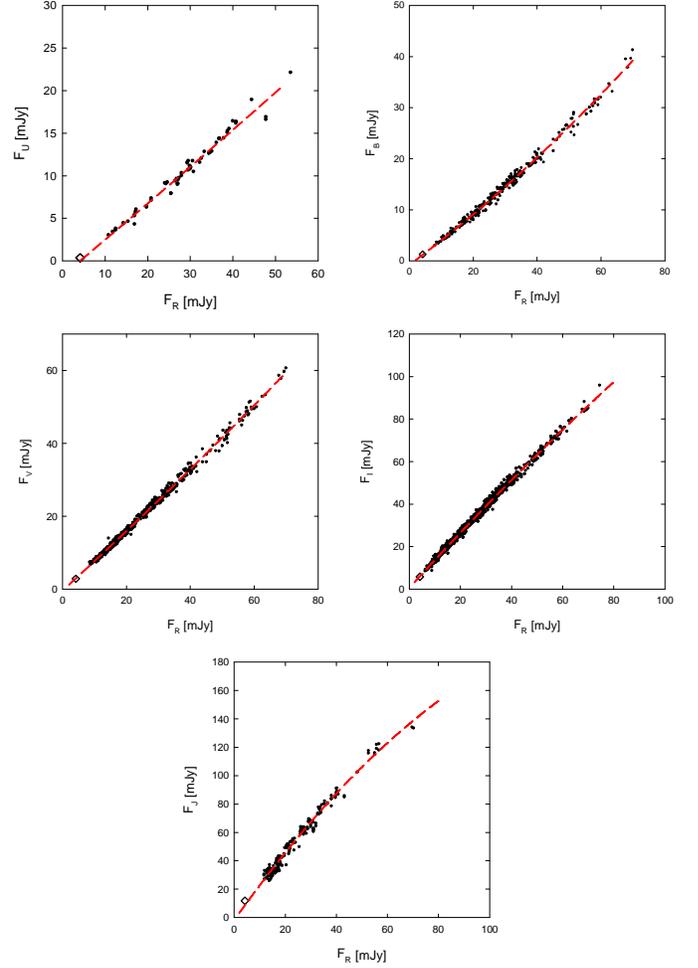}}
      \caption{De-reddened flux-flux dependences in the optical--NIR region. 
The red lines represent second-order polynomial regressions.
Diamonds correspond to the position of the underlying galaxy.}
         \label{bl_opt_regressions}
   \end{figure}

\begin{figure}
   \centering
  \resizebox{\hsize}{!}{\includegraphics[clip]{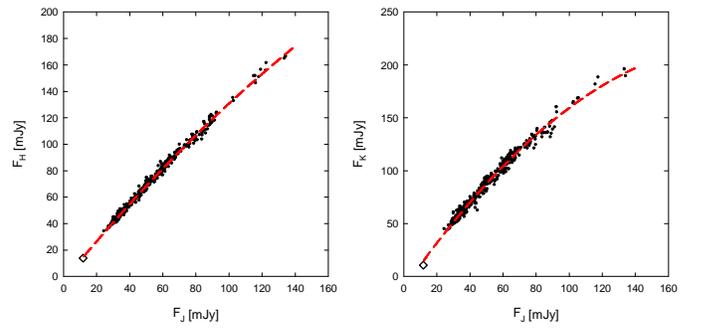}}
      \caption{Same as Fig.~\ref{bl_opt_regressions} for the NIR region.}
         \label{bl_opt_ir_regressions}
   \end{figure}

Figures~\ref{bl_opt_regressions} and~\ref{bl_opt_ir_regressions} display flux-flux dependences for optical and NIR bands in the best-sampled time interval 2000--2008. 
They show that in no case the relation between flux densities can be properly fitted by a linear dependence, but rather by second-order polynomials $a+b\cdot F_R+c\cdot F_R^2$ or $a+b\cdot F_J+c\cdot F_J^2$. 
The concavity of the regression is always directed toward the higher-frequency band, thus clearly displaying a bluer-when-brighter type of variable source behaviour. 
This result is in agreement with those obtained by \citet{Villata2002}, \citet{Villata2004a}, and \citet{Papadakis2007}, but it is in contrast with what found by \citet{Hagen-Thorn2004}. Indeed, the latter authors used a limited dataset, spanning a smaller flux range, so they could not identify deviations from a linear dependence in the flux-flux plots.

We assume that the deviations from regular variability seen in Figs.\ \ref{bl_opt_regressions} and~\ref{bl_opt_ir_regressions} are mostly due to noise. It cannot be excluded that some part of them are real, but their amplitude is small as compared to the overall variability pattern.

Now we need to take into account the host-galaxy contribution.
Its flux densities are obtained by adopting $R=15.55$ from \citet{Scarpa2000} and the mean colour indices for elliptical galaxies from \citet{Mannucci2001}. 
We get for the $UBVRIJHK$ galaxy flux densities: 0.36, 1.30, 2.89, 4.23, 5.90, 11.83, 13.97, and 10.62 mJy, respectively.
The flux positions of the host galaxy are marked in Figs.\ \ref{bl_opt_regressions} and~\ref{bl_opt_ir_regressions} with diamonds. It is remarkable that these positions lie almost exactly on the extrapolations of the second-order regression curves for the corresponding pairs of bands. This suggests that whatever mechanism of variability acts, most likely it is one and the same for the whole optical--NIR region.

All the measurements, according to WEBT prescriptions, were performed with a 8 arcsec aperture radius. Using a de~Vacouleurs' radial flux distribution and typical values of seeing, we estimate that $\approx 50\%$ of the galaxy flux density must be subtracted from the total photometric measure in order to get the flux density of the active nucleus.

\begin{figure}
   \centering
  \resizebox{\hsize}{!}{\includegraphics[clip]{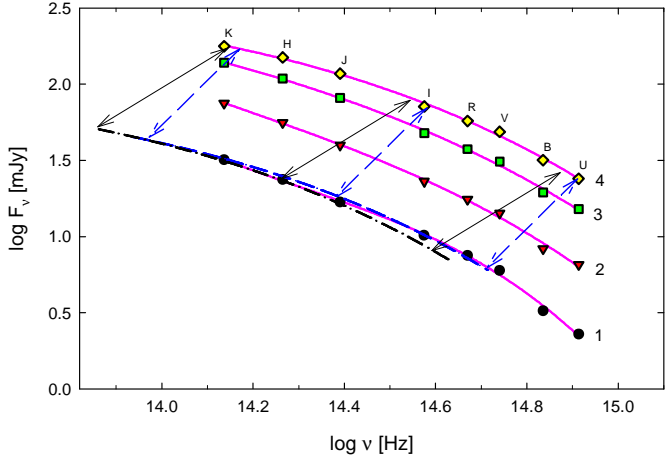}}
      \caption{Spectra of BL~Lacertae obtained by using the polynomial regressions 
shown in Figs.~\ref{bl_opt_regressions} and~\ref{bl_opt_ir_regressions} and subtracting the host galaxy contribution. 
 Numbers from 1 to 4 correspond to $R$-band flux densities of 10, 20, 40, and 60~mJy, respectively. 
 The pink lines tracing each spectrum represent cubic spline interpolations. 
 Blue dashed arrows show the direction and amount of the spectrum displacement when $\delta$ changes 
 by a factor 1.58 and $p=3$ in Eq.\ 1; grey arrows correspond to $p=2$ and a $\delta$ 
 variation of a factor 1.88.
 Blue dashed and black dot-dashed curves are the results of the shift of spectrum 4 toward spectrum 1
 in both cases.}
         \label{bl_low_high_SED3}
   \end{figure}

Fixing flux density values in $R$, the polynomial regressions of Figs.~\ref{bl_opt_regressions} and~\ref{bl_opt_ir_regressions} allow us to get the corresponding flux densities in all the other bands. After subtracting the host galaxy contribution in each band, we finally obtain simultaneous spectra, free from the host galaxy and chance fluctuations.
Figure~\ref{bl_low_high_SED3} shows spectra of the variable source in BL~Lacertae for four selected flux densities in $R$ from 10 to 60~mJy. 

\section{The origin of the variability}

In this section we investigate the possibility that the observed flux variations come from changes of the Doppler factor $\delta$, and that these are in turn due to viewing angle variations of the emitting region, as suggested by \citet{Villata2002}, \citet{Villata2004a}, and \citet{Villata2009a}.

We recall that 
\begin{equation}
\delta=[\Gamma(1-\beta \cdot \cos\theta)]^{-1}, 
\label{eq:delta}
\end{equation}
$\beta$ being the bulk velocity of the emitting plasma in units of the speed of light, 
$\Gamma=(1-\beta^2)^{-1/2}$ the corresponding Lorentz factor, 
and $\theta$ the angle between the velocity vector and the line of sight.

In the observer's rest frame, the flux density is a function of the Doppler factor $\delta$ \citep{Rybicki1979,Urry1995}

\begin{equation}
	F_\nu(\nu)=\delta^p\cdot F'_{\nu'}(\nu')
\label{eq:doppler1}
\end{equation}

\noindent where primed quantities refer to the source rest frame, and 
$p=3$ in case the radiation comes from a moving source (such as a discrete, essentially point-like, emitting zone), while $p=2$ for a smooth, continuous jet \citep{Begelman1984,Cawthorne1991}.

Doppler shift also affects frequencies:

\begin{equation}
	\nu=\delta\cdot\nu'.
\label{eq:doppler2}
\end{equation}

If we assume that the flux variations are due to changes in $\delta$, only
in the case when the intrinsic spectrum is a power law $F'_{\nu'} (\nu') \propto (\nu')^{-\alpha}$ and $\alpha$ is constant in the wavelength region under consideration, the flux-flux plots presented in the previous section would show linear dependences.
Indeed, only in this case Doppler boosting effects on both $F_\nu$ and $\nu$ lead to an unchanged observed spectral slope: 

\begin{equation}
F_\nu (\nu) \propto \delta^{p+\alpha} \cdot \nu^{-\alpha}.
\label{eq:flux}
\end{equation}
An example of that kind of behaviour was displayed by \object{3C~279} in 2006--2007 \citep{Larionov2008}.
However, in Sect.~2 we saw that the flux-flux relations are not linear, and indeed the spectra shown in Fig.~\ref{bl_low_high_SED3} are far from a power law.

As a consequence of Eqs.~\ref{eq:doppler1} and \ref{eq:doppler2}, in logarithmic scale (as in Fig.~\ref{bl_low_high_SED3})
changes of $\Delta \log\delta$ in the Doppler factor lead to a shift of the spectrum by $\Delta \log\nu=\Delta \log\delta$ 
(in the frequency direction) and $\Delta \log F_{\nu} = p\cdot\Delta \log\delta$ (in the flux direction).

The blue dashed spectrum in Fig.~\ref{bl_low_high_SED3} has been obtained from spectrum 4 by decreasing $\delta$ by a factor 1.58, and setting $p=3$. It fairly reproduces the ``real" spectrum 1. We cannot obtain the same agreement using $p=2$ (black dot-dashed curve). 
This suggests that the general variability pattern of BL~Lacertae during 2000--2008 in the optical and near-infrared part of the spectrum can be interpreted
in terms of Doppler factor variations affecting the radiation coming from a moving, discrete, and steady-emitting zone(s), possibly travelling in a jet.

Under the assumption that $\delta$ changes are due to viewing angle ($\theta$) variations,
we can try to estimate what $\theta$ variations are needed to account for the observed range of flux variability in the $R$ band.
From the definition of $\delta$ in Eq.~\ref{eq:delta}: 

\begin{equation}
\theta=\arccos{{\delta \, \Gamma - 1} \over {\delta \sqrt{\Gamma^2 - 1}}}.
\label{eq:phi} 
\end{equation}

With $p=3$, Eq.~\ref{eq:flux} becomes $F_{\nu}(\nu) = F_\circ \delta^{3+\alpha} \nu^{-\alpha}$. 
The mean value of $\alpha$ close to the $R$ band is $\alpha\approx1.3$ (see Fig.~\ref{bl_low_high_SED3}). The constant $F_\circ$ can be determined from $F_\circ=F_{\rm max}\nu^{\alpha}/\delta_{\rm max}^{3+\alpha}$, where $F_{\rm max}$ is the maximum observed flux density in $R$ band and $\delta_{\rm max}$ the corresponding maximum Doppler factor, which comes from the minimum viewing angle $\theta_{\rm min}$. We fix, as a tentative value, $\theta_{\rm min}=2\degr$. According to \citet{Marscher2008} and \citet{Jorstad2005}, $\Gamma=7.0\pm1.8$, then $\theta_{\rm min}=2\degr$ yields $\delta_{\rm max}=13.15$. 

\begin{figure}
   \centering
  \resizebox{\hsize}{!}{\includegraphics[clip]{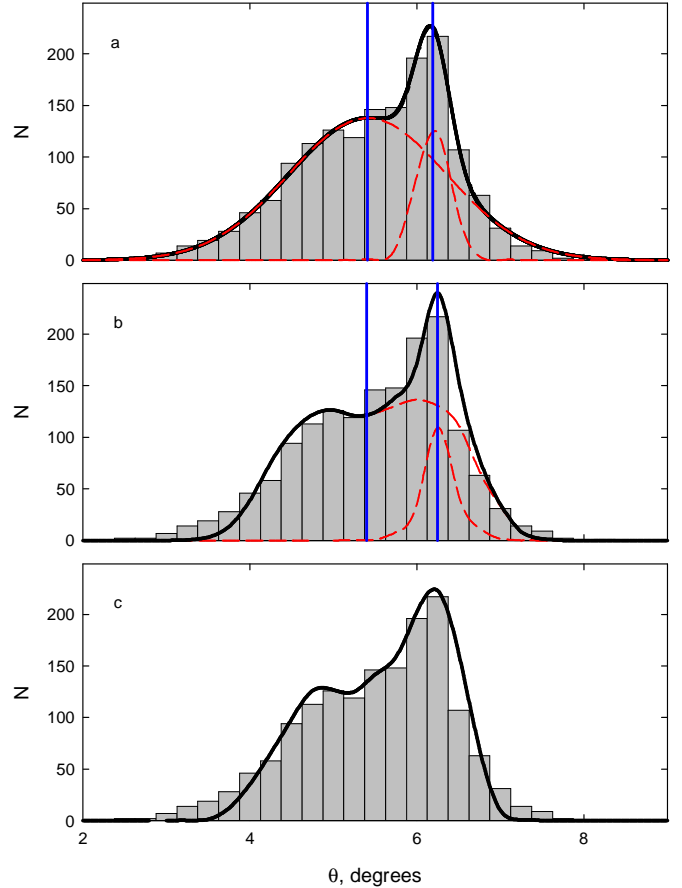}}
      \caption{Histogram of viewing angle $\theta$ distribution in 2000--2008, fitted with 
(a) two Gaussians; 
(b) ring-like Gaussian distribution plus Gaussian (red dashed curves indicate the contributions of the different components); 
(c) ring-like distribution with asymmetric density (see text).}
         \label{bl_histogram3}
   \end{figure}

Using these values in the above equations, we can get the pertaining values of $\theta$ for each observed $R$-band flux density. The histogram of the distribution of $\theta$ for all the time interval 
(2000--2008) considered is shown in Fig.~\ref{bl_histogram3}. In order to avoid false signals caused by uneven distribution of the data, we made nightly binning beforehand.

\begin{figure}
   \centering
\includegraphics[width=8cm,clip]{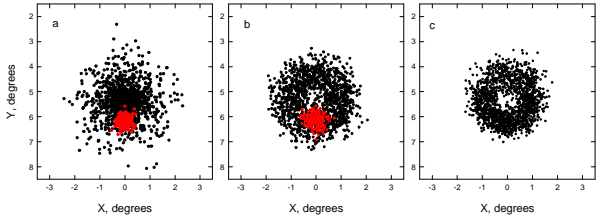} 
      \caption{Two-dimensional distributions of angles, corresponding to the three different interpretations 
of the histogram in Fig.~\ref{bl_histogram3}.}
         \label{bl_2d_distrib2}
   \end{figure}

Figure~\ref{bl_histogram3} shows that all the values of $\theta$ are grouped within the range from $2\degr$ to $8\degr$ (notice that the  value of $2\degr$ was fixed initially; setting another value would just shift all the histogram leftwards or rightwards). The resulting distribution shows substantial asymmetry and bimodality. 
Of course, the interpretation of these features is not unambiguous.  Red dashed lines in the two upper panels 
represent fitting with (a) two Gaussians and (b) ring-like structure plus Gaussian; blue vertical lines correspond to the centres of these components and the solid black lines indicate their sum. 
The bottom panel (c) represents the single-component case of a ring-like distribution with increasing density toward larger angles. These three model distributions are shown in Fig.~\ref{bl_2d_distrib2}.

In order to discriminate among them, we recall that we are working under the assumption that the flux variations of BL~Lacertae are mainly (if not solely) due to changes of the Doppler boosting. In this case it is necessary to take into account also time contraction, that leads to shortened time scales in the observer's rest frame: $t=\delta^{-1}t^\prime$. The degree of contraction is inversely proportional to the Doppler factor, which depends on the viewing angle of the velocity vector of the emitting source. It means that the smaller is $\theta$ during some interval in the light history of BL~Lac, the faster this interval will pass in the observer's rest frame. Quantitatively, in our case it implies that when $\theta$ changes from $8\degr$ to $2\degr$, the time scales shorten by a factor $\sim 2$. 

\begin{figure}
   \centering
\resizebox{\hsize}{!}
{\includegraphics[clip]{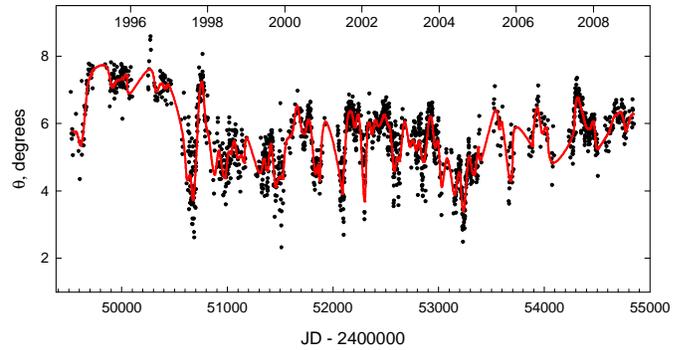}}
      \caption{Time history of the viewing angle $\theta$ from 1994 to 2008. 
The red line shows a cubic spline interpolation through the 30-day binned data.}
         \label{bl_phi_vs_time2}
   \end{figure}

This argument favours the model of Fig.~\ref{bl_2d_distrib2}c: the circular shape may reflect dominant spiral movement of emitting features \citep[see, e.g.][]{Marscher2008}, with time contraction leading to shorter life-time of high-flux-level events. This can also be seen in Fig.~\ref{bl_phi_vs_time2}, where we plot the time evolution of $\theta$ corresponding to the 1-day binned light curve in 1994--2008: smaller angles appear to last for a shorter time.

An alternative explanation of both Figs.~\ref{bl_2d_distrib2}c and \ref{bl_phi_vs_time2} can be found in the helical-jet model by \citet[][see also \citealt{Villata1999,Raiteri1999,Ostorero2004,Raiteri2009}]{Villata2009a}.
In this scenario, the changing-angle emitting zone is the whole optical--NIR emitting region in the helical jet, whose orientation changes because of the helix rotation. In this case Fig.~\ref{bl_2d_distrib2}c would represent all the 
viewing angles acquired by the emitting zone during this rotation in 2000--2008.
The angle time evolution in Fig.~\ref{bl_phi_vs_time2} is rather noisy and does not allow to recognise any clear periodicity, even if something like a yearly modulation can be envisaged (see also Fig.~1). 
The origin of the noise around the main modulation may reflect a more complex structure of the helical path,
such as a thinner helical structure of which the main helix would be the 
axis\footnote{This thinner helical structure may simply consist of helically twisted magnetic field lines,
whose toroidal component is needed for the MHD equilibrium of the jet \citep[see e.g.][and references therein]{Villata1995}.}. In this picture,
fast flares in the light curve (i.e.\ the noise in the angle curve and the thickness of the ring in Fig.~\ref{bl_2d_distrib2}c) could be due to shocks moving along the fine helical path and hence rapidly changing their
viewing angles. When these events occur during a minimum viewing angle orientation of the whole emitting region in the main helix, we observe the optical--NIR outbursts.

From Figs.~1 and 8 one can also see that before 1997 the optical activity was much lower.
According to \citet{Villata2004b}, \citet{Bach2006}, and \citet{Villata2009a} the 1994--1996 faint state shown
in the figures represents just the end of a longer quiescent period started in 1981.
\citet{Villata2009a} ascribed this low-moderate activity to a general bad alignment of the
optically emitting region with the line of sight. Namely, the axis of the main helix had a viewing angle larger
than that acquired in 1997. (Looking at the figures, it seems that this angle has been again increasing after 2004.)
According to this picture, if we had sufficient data in 1981--1996, we could compose a figure similar to Fig.~\ref{bl_2d_distrib2}c, but with the ring centre (representing the helix axis) farther from the line of sight.
Thus, Fig.~\ref{bl_2d_distrib2}c would be a necessary simplification that does not take the helix drift into account, and this may be the reason why the single ring does not fit well the ``wings" in Fig.~6c, which would belong to other, closer and farther, rings. 

\citet{Villata2009a} also argued that the helix axis is curved, to explain why the optical low
activity was actually accompanied by strong low-frequency radio activity, and vice versa \citep[see also][for the case of 3C 454.3]{Villata2007,Villata2009b}.
If the helix axis is curved, it may even be helical itself. 
In conclusion, we could be in the presence of a ``fractal" helical structure: the finest structure being responsible for fast flares (on day--weeks time scales), the intermediate one for months-lasting, roughly annual
outbursts, and the biggest structure for the longer-term (ten-yearly or more) alternation of low and high activity.

\section{Conclusions}

The analysis of the photometric $UBVRIJHK$ data collected during the 2000--2008 WEBT campaigns on BL~Lacertae, complemented by Crimean, St.\ Petersburg, and Campo Imperatore data in 2005--2006 and 2008, led us to the following conclusions:

\begin{enumerate}
\item{The distribution of the optical and NIR data on flux-flux plots indicates the presence of a single dominant source of variability with a bluer-when-brighter behaviour.}
\item{The dependence of the spectral shape on the flux level agrees with the suggestion that the variability in that spectral region is mostly caused by changes of the Doppler factor.}
\item{The emission would likely come from a moving, discrete, and steady-emitting zone(s), possibly travelling in a jet.}
\item{Under the assumption that the Doppler factor variations are due to changes of the viewing angle, the histogram of viewing angle distribution appears asymmetric and can be fitted with a ring-shaped pattern with uneven angular distribution. We infer that this is caused by time contraction in the observer's rest frame, which sharpens the high-flux (small viewing angle) events.}
\item{The noise around the main viewing angle modulation could be the signature of a finer jet structure. The above mentioned discrete emitting zones (e.g.\ shock waves) would travel along 
a short-pitch helical path, thus rapidly changing their viewing angle to produce fast flares.
The axis of this helix would be helical itself, and its changing viewing angle due to rotation would cause the roughly annual outbursts. 
This main helix, in turn, would have a curved axis, maybe helical again, whose orientation changes would yield the observed long-term alternation of high and low optical activity.
In summary, the jet might exhibit a fractal helical structure, where increasing geometric scales correspond to increasing variability time scales.}
\end{enumerate}

The analysis presented in this paper is based on data covering a limited energy range (optical--NIR). However, our results confirm the geometrical interpretation of BL Lacertae variability suggested in previous works based on WEBT data, where broader-band observations were available. We cannot rule out other explanations for the observed spectral variability, based on intrinsic rather than geometric mechanisms. However, we think that straight and motionless jets are not realistic (as the bent trajectories of jet components revealed by VLBI observations suggest), and thus changes in the viewing angle of the emitting jet regions are very likely to occur.
We expect that in the near future broad-band multiwavelength campaigns, including $\gamma$-ray observations by the Fermi-GST and AGILE satellites, will help clarify the picture on blazar variability.
Indeed, they will allow people to study simultaneously both the low-energy synchrotron and the high-energy inverse-Compton emission components, and thus to better test theoretical models.

\begin{acknowledgements}
We thank Alan P.\ Marscher and Svetlana G.\ Jorstad for useful discussions. V.L. acknowledges support from Russian RFBR foundation via grant 09-02-00092.
\end{acknowledgements}

\end{document}